\documentclass[11pt]{article}
\usepackage{axodraw}
\usepackage{epsfig}
\usepackage{amsfonts}
\usepackage{amsmath}
\usepackage{bbm,bm}
\allowdisplaybreaks[1]

\input pix.sty

\newcommand{\ko}{k_0}
\newcommand{\km}{k_-}
\newcommand{\kp}{k_+}

\renewcommand{\eq}{eq.~}
\renewcommand{\eqs}{eqs.~}
\renewcommand{\se}{sec.~}

\renewcommand{\fig}{fig.~}

\newcommand{\tinymsbar}{{\overline{\mbox{\tiny\rm{MS}}}}}
\newcommand{\Lambdamsbar}{{\Lambda_\tinymsbar}}

\newcommand{\alphas}{\alpha_{\rm s}}
\newcommand{\Nf}{N_{\rm f}}
\newcommand{\Nc}{N_{\rm c}}

\newcommand{\CF}{C_\rmii{F}}

\newcommand{\mE}{m_\rmii{E}}
\newcommand{\gE}{g_\rmii{E}}
\newcommand{\gammaE}{\gamma_\rmii{E}}

\newcommand{\rmO}{{\mathcal{O}}}
\newcommand{\bmu}{\bar\mu}

\def\lsi{\raise0.3ex\hbox{$<$\kern-0.75em\raise-1.1ex\hbox{$\sim$}}}
\def\gsi{\raise0.3ex\hbox{$>$\kern-0.75em\raise-1.1ex\hbox{$\sim$}}}
\newcommand{\lsim}{\mathop{\lsi}}
\newcommand{\gsim}{\mathop{\gsi}}

\newcommand{\nF}{n_\rmii{F}}
\newcommand{\nB}{n_\rmii{B}}
 \renewcommand{\nF}[1]{n_\rmii{F{#1}}}
 \renewcommand{\nB}[1]{n_\rmii{B{#1}}}

\newcommand{\rmii}[1]{{\mbox{\tiny\rm{#1}}}}

\newcommand{\re}{\mathop{\mbox{Re}}}
\newcommand{\im}{\mathop{\mbox{Im}}}

\newcommand{\Tint}[1]{{\hbox{$\sum$}\!\!\!\!\!\!\!\int\,}_{\!\!\!\!\raise-0.9ex\hbox{$\scriptstyle{#1}$}}}
\newcommand{\Tinti}[1]{{{\Sigma}\!\!\!\!\raise0.3ex\hbox{$\int$}_\rmii{${#1}$}}}

\newcommand{\bi}{\begin{itemize}}
\newcommand{\ei}{\end{itemize}}

\newcommand{\hide}[1]{ }

\def\TAsc(#1,#2)(#3,#4,#5)%
{\SetWidth{2.0}\CArc(#1,#2)(#3,#4,#5)\SetWidth{1.0}}
\def\Lwidth{3}

\def\TAgl(#1,#2)(#3,#4,#5){\SetWidth{2.0}\PhotonArc(#1,#2)(#3,#4,#5){\Lwidth}%
{6.283 #3 mul 360 div #4 #5 sub #4 #5 sub mul sqrt mul Tdensity mul}%
\SetWidth{1.0}}
\def\TLgl(#1,#2)(#3,#4){\SetWidth{2.0}\Photon(#1,#2)(#3,#4){\Lwidth}
{#1 #3 sub #1 #3 sub mul #2 #4 sub #2 #4 sub mul add sqrt Tdensity mul}%
\SetWidth{1.0}}

\renewcommand{\picc}[1]{\;\parbox[c]{60pt}{\begin{picture}(60,30)(0,0)
\SetWidth{1.0}\SetScale{1.3} #1 \end{picture}}\; }
\def\Lwidth{1.3}

\makeatletter \@addtoreset{equation}{section} \makeatother
\renewcommand{\theequation}{\arabic{section}.\arabic{equation}}
\makeatletter
\renewcommand\section{\@startsection {section}{1}{\z@}%
                                   {-5.5ex \@plus -1ex \@minus -.2ex}
                                   {2.3ex \@plus.2ex}%
                                   {\normalfont\large\bfseries}}
\renewcommand\subsection{\@startsection{subsection}{2}{\z@}%
                                     {-3.25ex\@plus -1ex \@minus -.2ex}%
                                     {1.5ex \@plus .2ex}%
                                     {\normalfont\normalsize\bfseries}}
\renewcommand\thesection {\@arabic\c@section}
\renewcommand\thesubsection   {\thesection.\@arabic\c@subsection}
\renewcommand{\@seccntformat}[1]{%
\csname the#1\endcsname.\hspace{1.0em}}
\makeatother

\begin{document}

\flushbottom

\begin{titlepage}

\begin{flushright}
\vspace*{1cm}
\end{flushright}
\begin{centering}
\vfill

{\Large{\bf
 Interpolation of hard and soft dilepton rates
}} 

\vspace{0.8cm}

I.~Ghisoiu and M.~Laine 

\vspace{0.8cm}

{\em
Institute for Theoretical Physics, 
Albert Einstein Center, University of Bern, \\ 
Sidlerstrasse 5, CH-3012 Bern, Switzerland\\}

\vspace*{0.8cm}

\mbox{\bf Abstract}
 
\end{centering}

\vspace*{0.3cm}
 
\noindent
Strict next-to-leading order (NLO) results for the dilepton production rate
from a QCD plasma at temperatures above a few hundred MeV suffer from 
a breakdown of the loop expansion in the regime of soft invariant masses 
$M^2 \ll (\pi T)^2$. In this regime an LPM resummation is needed for
obtaining the correct leading-order result. We show how to construct 
an interpolation between the hard NLO and the leading-order LPM expression. 
Numerical results are presented in a tabulated 
form, suitable for insertion into hydrodynamical codes. 

\vfill

 
\vspace*{1cm}
  
\noindent
September 2014

\vfill

\end{titlepage}

%
\section{Introduction}
\la{se:intro}

Consider $\mu^-\mu^+$ or $e^-e^+$ pairs produced thermally from 
a quark-gluon plasma at a temperature $T \gsim 150$~MeV, with the pair having
a non-zero total momentum $k \equiv |\vec{k}| \sim $~GeV
with respect to the plasma rest frame, and 
an invariant mass $M \sim $~GeV. If no zero-temperature 
resonance lies near the $M$ considered, 
non-thermal backgrounds for the production of such dileptons
are expected to be smaller than for on-shell photons, and  
dileptons may therefore constitute a good probe of
QCD interactions at finite temperature. As a particular reflection
of deconfinement and chiral symmetry restoration, 
a relatively smooth shape is anticipated for the thermal
dilepton production rate, with a characteristic overall 
magnitude (to be determined by theoretical computations)
and an exponentially damped spectral shape.  

Many different approximation schemes and kinematic regimes
have been considered for thermal dilepton production in the past. First 
next-to-leading order (NLO) analyses were carried out long ago 
for $k =0$~\cite{old1,old2,old3}, finding that for $M\sim \pi T$
radiative corrections are in general small. However, moving to a ``soft'' 
invariant mass
$M\sim gT$ with still $k=0$ (here $g^2 \equiv 4\pi\alphas$), 
a major enhancement of the rate
was found after carrying out Hard Thermal Loop (HTL) resummation~\cite{bp}.
Most of the past work has concentrated on $M \sim gT$
but large spatial momentum ($k \gsim \pi T$). In this regime
the NLO rate has a logarithmic singularity, which is regulated
by Landau damping of the quarks mediating $t$-channel 
exchange~\cite{kapusta,baier}.
In addition, there are finite terms which all contribute 
at the same order because of collinear enhancement, 
and need to be handled through Landau-Pomeranchuk-Migdal (LPM) 
resummation~\cite{agz_m} (LPM resummation incorporates HTL resummation
in an approximation valid for $k\gg gT$).
In order to avoid double counting, LPM resummation 
needs to be carefully combined with other processes~\cite{cg}.
A resummation beyond HTL (based on effective kinetic theory) 
is also needed at $k=0$ for $M \ll gT$~\cite{mr}.
In contrast, for $M \gg \pi T$ no resummation is needed
at NLO, and the analysis can 
be greatly simplified by making use of Operator Product Expansion (OPE) 
techniques, with the results available in analytic
form~\cite{sch}. Unfortunately the OPE expansion shows convergence
only quite deep in the hard regime ($M \gg 8T$)~\cite{dilepton}.
Finally, lattice simulations are being carried out at
$k=0$~\cite{hengtong1,mainz,swansea} and  
at $k\neq 0$~\cite{hengtong2}, even though the usual issues
with analytic continuation imply that the results may suffer
from uncontrolled systematic uncertainties~\cite{cond}. 

As is clear from the previous paragraph, many computations 
have concentrated on special regimes in which one or 
the other kinematic simplification can be made. The current
study is a continuation of a recent project~\cite{master,relat} 
which led to the determination of the NLO dilepton rate for {\em generic}
momenta and invariant masses $k,M \sim \pi T$~\cite{dilepton}. 
The goal of the present study is to present a smooth interpolation between 
these hard NLO expressions, and leading-order LPM resummation in the soft 
regime $M\ll \pi T$, $k \gg M$. The interpolated results turn out 
to have a qualitatively correct behaviour even when
extrapolated down to $M\ll \pi T$, $k \sim 0$. Therefore, for 
practical purposes, we hope that they yield a fair estimate
of the thermal dilepton rate from a deconfined QCD plasma
for the invariant masses and spatial
momenta of interest to the current heavy ion collision program. 
 
The plan of this paper is the following. 
After defining the observables to be considered in \se\ref{se:setup},
we briefly review the status of hard NLO computations in  
\se\ref{se:hard} and of soft LPM resummation in \se\ref{se:lpm}
(we also introduce an efficient method for the numerical 
solution of the LPM equations). 
A way to consistently combine these approaches is 
explained in \se\ref{se:combination}. Numerical results, 
meant for phenomenological use, are displayed in \se\ref{se:spectra}, 
and we finish with a brief conclusion and outlook in \se\ref{se:concl}.

%
\section{Basic definitions}
\la{se:setup}

To leading order in 
$\alpha^{ }_{e} \equiv e^2/(4\pi)$~\cite{dilepton1,dilepton2,dilepton3} 
and omitting power-suppressed corrections from 
$Z$-boson exchange, the production rate 
of $\mu^-\mu^+$ pairs from a hot QCD medium, 
with a total four-momentum 
$
 \mathcal{K} \equiv  \mathcal{K}^{ }_{\mu^-} + \mathcal{K}^{ }_{\mu^+} 
 \equiv (\ko,\vec{k})
$, 
can be expressed as
\ba
 \frac{{\rm d} N_{\mu^-\mu^+}}
   {{\rm d}^4 \mathcal{X} {\rm d}^4 \mathcal{K}} 
  & {=} &   
 - \frac{  \alpha_{e}^2  } 
  {3 \pi^3 \mathcal{K}^2} 
 \, \biggl( 
   1 + \frac{2 m_\mu^2}{\mathcal{K}^2}
 \biggr)
 \, \biggl(
   1 - \frac{4 m_\mu^2}{\mathcal{K}^2} 
 \biggr)^\fr12 
 \, \theta(\mathcal{K}^2 - 4 m_\mu^2)
 \, \nB{} (\ko) 
  \nn & \times & 
 \biggl[\Bigl( 
 \sum_{i = 1}^{\Nf} Q_i^2\Bigr) \;  
 \rho_\rmii{NS}^{ }(\mathcal{K}) + 
 \Bigl( \sum_{i = 1}^{\Nf} Q_i \Bigr)^2 \; 
 \rho_\rmii{SI}^{ }(\mathcal{K}) 
 \biggr]
 \;. 
 \la{physics} 
\ea
Here $\nB{}$ is the Bose distribution, and 
$\rho^{ }_\rmii{NS}$ and $\rho^{ }_\rmii{SI}$ denote
spectral functions in the ``non-singlet''
and ``singlet'' channels, respectively, with the quark flavours
assumed degenerate for simplicity. 
For $\Nf = 3$ the singlet channel drops out, 
and we concentrate on 
\ba
 \rho_\rmii{NS}(\mathcal{K}) &  \equiv &
  \int_\mathcal{X} 
   e^{i \mathcal{K}\cdot \mathcal{X}}
  \left\langle
    \fr12 \bigl[ 
    \hat{\mathcal{J}}^\mu (\mathcal{X}), 
    \hat{\mathcal{J}}^{ }_{\mu}(0)
    \bigr]
  \right\rangle^{ }_{\rmi{c}}
  \;, \quad
    \hat{\mathcal{J}}^\mu \equiv  \hat{\bar{\psi}}\gamma^\mu\hat{\psi}
  \;, \la{rho_NS}
\ea 
where $c$ denotes a connected quark contraction; 
$\eta_{\mu\nu} \equiv \mbox{diag}(+$$-$$-$$-)$; 
and $\int_\mathcal{X}$ is an integral over the spacetime volume. 
According to \eq\nr{physics}, 
$\rho^{ }_\rmii{NS}$ must be negative, 
so we mostly discuss
\be
 - \rho^{ }_\rmii{NS}(\mathcal{K})
    \; =  \;
 - \im \Pi^{ }_\rmii{R}(\mathcal{K})
    \; > \; 0
 \la{relation}
\ee
in the following, 
where $\Pi^{ }_\rmii{R}$ refers to the retarded correlator.  

Let us inspect
separately the ``transverse'' and ``longitudinal'' parts of 
$ - \im \Pi^{ }_\rmii{R} $. Choosing for convenience the $z$-axis
to point in the direction of $\vec{k}$, 
\be
 \vec{k} \equiv (0,0,k)
 \;,
\ee
the transverse part is 
\be
 - \im \Pi^{ }_\rmii{R,T} \; \equiv \; 
  \sum_{i=1}^{2}
  \int_\mathcal{X} 
   e^{i \mathcal{K}\cdot \mathcal{X}}
  \left\langle
    \fr12 \bigl[ 
    \hat{\mathcal{J}}^{ }_{i} (\mathcal{X}), 
    \hat{\mathcal{J}}^{ }_{i}(0)
    \bigr]
  \right\rangle^{ }_{\rmi{c}}
 \;.  \la{imRT}
\ee
The remaining longitudinal part can be expressed as 
\be
  - \im \Pi^{ }_\rmii{R,L} \; \equiv \; 
  \im \Pi^{ }_\rmii{R,33} - \im \Pi^{ }_\rmii{R,00}  
  \; = \; 
  \frac{\mathcal{K}^2}{k^2} \, 
  \int_\mathcal{X} 
   e^{i \mathcal{K}\cdot \mathcal{X}}
  \left\langle
    \fr12 \bigl[ 
    \hat{\mathcal{J}}^{ }_{0} (\mathcal{X}), 
    \hat{\mathcal{J}}^{ }_{0}(0)
    \bigr]
  \right\rangle^{ }_{\rmi{c}}
  \;, \la{imRL}
\ee
where we made use of a Ward identity relating
$\im \Pi^{ }_\rmii{R,00}$ 
and 
$\im \Pi^{ }_\rmii{R,33}$.
The physically relevant combination is 
\be
  - \im \Pi^{ }_\rmii{R}(\mathcal{K}) = 
  - \im \Pi^{ }_\rmii{R,T}(\mathcal{K}) 
  - \im \Pi^{ }_\rmii{R,L}(\mathcal{K})
 \;. \la{obs}
\ee

%
\section{NLO dilepton rate for general momenta}
\la{se:hard}

The observable of \eq\nr{obs} (although not separately its two parts
$\im \Pi^{ }_\rmii{R,T}$, 
$\im \Pi^{ }_\rmii{R,L}$)
is currently known up to NLO in a strict loop expansion~\cite{dilepton}. 
However only the leading-order (LO) result can be given in analytic form: 
\ba
 - \left. \im \Pi_\rmii{R}(\mathcal{K}) \right|_{ }^{(g^0)}
 & = & \frac{ \Nc T M^2 }{2\pi k }
 \; 
 \ln\biggl\{
   \frac{\cosh\bigl(\frac{\kp}{2 T} \bigr) }
        {\cosh\bigl(\frac{\km}{2 T} \bigr) }  
 \biggr\}
 \;. \la{pert_LO}
\ea
Here light-cone momenta and a photon invariant mass were defined as
\be
 k^{ }_\pm \equiv \frac{\ko \pm k}{2} > 0 
 \;, \quad
 M \equiv \sqrt{\mathcal{K}^2}
 \;. \la{kpm}
\ee 

For future reference, it is helpful
to present \eq\nr{pert_LO} 
also in a form before a final integration. We do this 
separately for the parts in \eqs\nr{imRT} and \nr{imRL}: 
\ba
 - \left. \im\Pi^\rmii{ }_\rmii{R,L}(\mathcal{K}) \right|_{ }^{(g^0)} & = & 
 \frac{4 \Nc M^2}{k^2}
 \biggl[
    \frac{ k^2 - \ko^2 }{2} \bigl\langle 1 \bigr\rangle 
  + 2 \bigl\langle \omega(\ko - \omega) \bigr\rangle  
 \biggr]
 \;, \la{rhoL}
 \\ 
 - \left. \im\Pi^\rmii{ }_\rmii{R,T}(\mathcal{K}) \right|_{ }^{(g^0)} & = & 
 \frac{4 \Nc M^2}{k^2}
 \biggl[
    \frac{ k^2 + \ko^2 }{2} \bigl\langle 1 \bigr\rangle 
  - 2 \bigl\langle \omega(\ko - \omega) \bigr\rangle  
 \biggr]
 \;, \la{rhoT}
\ea
where 
\ba
 \langle ... \rangle 
 & \equiv &  \frac{1}{16\pi k}
 \int_{\km}^{\kp} \! {\rm d}\omega \, 
 \bigl[
   1  - \nF{}(\omega) - \nF{}(\ko - \omega)   
 \bigr] \, (...) 
 \;. \la{ave}
\ea
It is seen that a substantial cancellation takes place 
when adding up \eqs\nr{rhoL}, \nr{rhoT}.

At NLO, it is more cumbersome to 
work out analytic expressions. However 
a convergent 2-dimensional integral 
representation can be given~\cite{dilepton}.
An analytic result is obtained on one hand 
for the dominant logarithmic
divergence at $M\ll \pi T$~\cite{pvr}, 
and on the other for $M \gg \pi T$~\cite{sch}. 
Let us define an ``asymptotic'' thermal quark mass by 
\be
 m_\infty^2 
  \;\equiv\;  2 g^2 \CF
  \int_\vec{p} \frac{\nB{}(p) + \nF{}(p)}{p}
  = \frac{g^2 \CF T^2}{4}
 \;,
\ee
where $\nF{}$ is the Fermi distribution, 
$p \equiv |\vec{p}|$, and
$
 \int_\vec{p} \equiv \int \!{\rm d}^3\vec{p}/(2\pi)^3
$.
Then the soft divergence reads
\be
 - \left. \im \Pi^{ }_\rmii{R}(\mathcal{K}) \right|_{ }^{(g^2)}
  \; \stackrel{M \,\ll\, \pi T}{\approx} \;
  \frac{\Nc m_\infty^2}{4\pi} \ln \Bigl( \frac{T^2}{M^2} \Bigr)
  \Bigl[ 1 - 2 \nF{}(\ko) \Bigr]
  + \rmO(\alphas T^2)
  \;, \la{log_div}
\ee
whereas the asymptotic expansion in the hard limit is given by 
\be
 - \im\Pi^{ }_\rmii{R}(\mathcal{K}) 
 \; \stackrel{M \,\gg\, \pi T}{\approx} \;
 \frac{\Nc {M}^2}{4\pi} 
 \left(\! 1 + \frac{3 \alphas \CF}{4\pi} \!\right)  
 + 
 \frac{4 \alphas \Nc \CF}{9}
 \left(\! 1 + \frac{4 k^2}{3 M^2} \!\right)
 \frac{\pi^2 T^4}{M^2}
 + \rmO\Bigl( \frac{\alphas T^6}{M^4} \Bigr)
 \;. \la{ope}
\ee

%
\section{LPM resummation near the light cone}
\la{se:lpm}

%
\subsection{Basic equations}
\la{ss:basic}

Leading-order Landau-Pomeranchuk-Migdal (LPM) resummation for
the dilepton production rate was worked out in ref.~\cite{agz_m}. 
The dilepton case is a generalization of the 
on-shell photon production rate that had been considered 
previously~\cite{amy1,amy2}. A field-theoretic derivation
of the basic equations can be found in ref.~\cite{bb}, and yet
another approach yielding the same dynamics, operating within
the imaginary-time formalism, in ref.~\cite{screening}.

In its usual formulation  LPM resummation
assumes the kinematics $\ko \gg g T$ and $\ko - k \ll \ko$.
Then only the 
leading terms in a Taylor expansion around 
the light cone $\ko=k$ are relevant. 
Parametrizing the kinematics 
through $\ko$ and $M^2$, this means that the spatial momentum $k$ 
can be expressed as 
\be
 k = \ko - \frac{M^2}{2\ko} \;,
 \la{k_appro} 
\ee
and the validity of this expansion is assumed in all formal manipulations
of the present section. (Numerically, however, we may at times exit the 
regime in which \eq\nr{k_appro} is literally accurate; the procedure to 
be followed in these cases is discussed below.)

Because of the assumption $\ko,k\gg gT$, 
Hard Thermal Loop (HTL) self-energies 
and vertices can be simplified through a 
``hard-particle'' approximation
(cf.\ e.g.\ ref.~\cite{lpm15}), resulting in an effective
kinetic description~\cite{eft} with particles carrying ``asymptotic''
thermal masses~\cite{gT}. 
With a minor change of conventions with respect to ref.~\cite{agz_m} 
(reshuffling of imaginary units; 
inversion of the sign of one of the frequency variables appearing; 
rescaling of wave functions;  
and use of $\nF{}(-\omega) = 1 - \nF{}(\omega)$), 
we are then led to define a ``2-particle Hamiltonian'',
\be
 \hat{H} \equiv 
 - \frac{M^2}{ 2\ko } 
 + \Bigl( \frac{1}{2 \omega_1} + \frac{1}{2 \omega_2} \Bigr) 
   \Bigl( m_\infty^2 - \nabla_\perp^2 \Bigr)
 + i V^+
 \;, \la{H} 
\ee
where $\nabla^{ }_\perp$ operates 
in the two ``transverse'' directions.\footnote{%
 The sign of the imaginary part is a convention; it could be reversed
 by a corresponding sign change on the right-hand sides of 
 \eqs\nr{imPiL_lpm} and \nr{imPiT_lpm}.
 }  
The light-cone potential is~\cite{agz,sch} 
\be
 V^+  =  \frac{ \gE^2 \CF }{2\pi} 
  \biggl[ 
    \ln\Bigl( \frac{\mE y }{2} \Bigr) + \gammaE + K_0^{ }(\mE y )
  \biggr] + \rmO\Bigl( \frac{\gE^4}{\mE} \Bigr) 
 \;, \la{V} 
\ee
where $y \equiv |\vec{y}|$ denotes a 2-dimensional transverse separation; 
$\CF \equiv (\Nc^2 - 1)/(2\Nc)$; 
$\gE^2 = g^2 T$ is the gauge coupling of the EQCD effective theory;
$\mE^2 = (\frac{\Nc}{3} + \frac{\Nf}{6})g^2 T^2$ is an electric mass
parameter in EQCD; and $K_0$ is a modified Bessel function. The superscript
in $V^+$ is a reminder of the fact that the potential is positive
for all $y > 0$ (it vanishes for $y=0$).

We need to solve Schr\"odinger equations in the $S$ and $P$-wave channels:
\ba
 \bigl( \hat{H} + i 0^+ \bigr) g(\vec{y})
 & = & \delta^{(2)}(\vec{y}) 
 \;, \la{g_eq} \\
 \bigl( \hat{H} + i 0^+ \bigr) \vec{f}(\vec{y})
 & = & -\nabla_\perp \delta^{(2)}(\vec{y}) 
 \;. \la{f_eq} 
\ea  
Then the functions we are interested in are 
\ba
  \im\Pi^\rmii{ }_\rmii{R,L} & = & 
  \Nc \int_{-\infty}^{\infty} \! {\rm d} \omega_1 
  \int_{-\infty}^{\infty} \! {\rm d} \omega_2 \, 
  \delta(\ko - \omega_1 - \omega_2) 
  \bigl[1 - \nF{}(\omega_1) - \nF{}(\omega_2) \bigr]
  \nn & & \, \times \, 
  \frac{M^2}{\ko^2}
  \lim_{\vec{y}\to\vec{0}}
  \frac{\im[g(\vec{y})]}{\pi}
 \;, \la{imPiL_lpm} \\
  \im\Pi^\rmii{ }_\rmii{R,T} & = & 
  \Nc \int_{-\infty}^{\infty} \! {\rm d} \omega_1 
  \int_{-\infty}^{\infty} \! {\rm d} \omega_2 \, 
  \delta(\ko - \omega_1 - \omega_2) 
  \bigl[1 - \nF{}(\omega_1) - \nF{}(\omega_2) \bigr]
  \nn & & \, \times \, 
  \biggl( \frac{1}{2\omega_1^2} + \frac{1}{2\omega_2^2} \biggr)
  \lim_{\vec{y}\to\vec{0}}
  \frac{\im[\nabla_\perp \cdot \vec{f}(\vec{y})]}{\pi}
 \;. \la{imPiT_lpm}
\ea
The Dirac-$\delta$ constraints here correspond to energy conservation, 
whereas the Schr\"odinger equations in \eqs\nr{g_eq}, \nr{f_eq} can be 
viewed as reflecting momentum conservation. 

Of the variables characterizing external kinematics ($\ko,k,M$), 
only two are independent ($\ko^2 - k^2 = M^2$).
As mentioned above, the derivation of the LPM equations 
can be justified as a leading term in 
a Taylor expansion in $M^2/k^2$ for $k \gg g T$. This implies
that $\ko,k,M$ should be related through \eq\nr{k_appro}. Sometimes, 
it may however be convenient to also apply the LPM equations beyond
their parametric validity range. There is no unique way of doing this, 
however one possible criterion 
is that there be a specific cancellation between the transverse and 
longitudinal contributions, namely that the terms 
$\langle \omega(\ko - \omega) \rangle$ in \eqs\nr{rhoL}, \nr{rhoT}
drop out. In order 
to maintain this cancellation within the LPM equations, the
coefficient $M^2/\ko^2$ in \eq\nr{imPiL_lpm} needs to be 
related to the variables appearing in \eq\nr{H} in a specific way.
This means that we have to give up either the strict 
$M^2/k^2$ multiplying $\im \Pi^{ }_\rmii{R,00}$
in \eq\nr{imRL} or the strict $k-\ko$ that is represented by 
$-M^2/(2\ko)$ in \eq\nr{H}. We have adopted a procedure, 
corresponding to ref.~\cite{agz_m}, where a compromise
has been made at both points; 
however we have verified that the numerical effect of other choices, 
if made consistently, is small. 

Following ref.~\cite{agz_m}, it is helpful for the following
to define a parameter $M_\rmi{eff}^2$
originating from a combination identifiable in \eq\nr{H}:  
\ba
 \Bigl( \frac{1}{2 \omega_1} + \frac{1}{2 \omega_2} \Bigr) M_\rmi{eff}^2
 & \equiv & 
 \left \{  
 - \frac{M^2}{2\ko}
 + \Bigl( \frac{1}{2 \omega_1} + \frac{1}{2 \omega_2} \Bigr) m_\infty^2 
 \right\} ^{ }_{ \omega_1 + \omega_2 = \ko }
 \;, \quad \\
 M_\rmi{eff}^2 & = & m_\infty^2 - 
 \frac{\omega_1 \omega_2}{\ko^2} M^2
 \;. \la{Meff}
\ea

%
\subsection{Method for numerical solution}
\la{num}

In order to solve \eqs\nr{g_eq}, \nr{f_eq} numerically, 
we adapt to two dimensions a
method employed in appendix~A of 
ref.~\cite{peskin} for solving vector and scalar 
channel quarkonium spectral functions in three dimensions. The basic
approach was introduced in ref.~\cite{original} for the vector channel
($S$-wave) case at zero temperature.
Its idea is to reduce the solution of an 
inhomogeneous equation to determining that solution of the homogeneous
equation which is regular at origin. 

By rescaling the transverse variable as $\rho \equiv y \mE $; introducing
a coordinate $\bm{\rho}'$ as a handle on the behaviour of the 
solution under rotations; rescaling the wave
functions into a dimensionless form; 
and making use of the parameter $M^2_\rmi{eff}$
introduced in \eq\nr{Meff},  
the inhomogeneous Schr\"odinger equations in \eqs\nr{g_eq}, \nr{f_eq}
can be re-expressed as specific limits of  
\be
 \biggl\{
   \frac{M_\rmii{eff}^2}{\mE^2} - \nabla_{\bm{\rho}}^2 + i \, 
 \biggl[ 
   \frac{2\omega_1\omega_2 V^+(\rho)}{\ko\mE^2} 
 \biggr] 
 \biggr\} \, \phi(\bm{\rho},\bm{\rho}') = 
 \delta^{(2)}(\bm{\rho}- \bm{\rho}') 
 \;. \la{full_S_eq}
\ee
In these variables 
the structures needed in \eqs\nr{imPiL_lpm}, \nr{imPiT_lpm} read
\ba
 \lim_{\vec{y}\to\vec{0}}
  \frac{\im[g(\vec{y})]}{\pi} & = & 
 \lim_{\bm{\rho},\bm{\rho}'\to\vec{0}} 
  \frac{2\omega_1\omega_2}{\pi \ko}  
 \im \bigl[ \phi(\bm{\rho},\bm{\rho}') \bigr]
 \;, \\ 
  \lim_{\vec{y}\to\vec{0}}
  \frac{\im[\nabla_\perp \cdot \vec{f}(\vec{y})]}{\pi}
 & = & 
 \lim_{\bm{\rho},\bm{\rho}'\to\vec{0}} 
  \frac{2\omega_1\omega_2\mE^2 }{\pi \ko}  
 \im \bigl[
    \nabla_{\bm{\rho}} \cdot 
    \nabla_{\bm{\rho}'} 
    \, \phi(\bm{\rho},\bm{\rho}') \bigr]
 \;. 
\ea
In polar coordinates, $\bm{\rho} = (\rho,\phi )$, 
the solution of the corresponding homogeneous equation, 
\be
 \biggl\{
   \frac{M_\rmii{eff}^2}{\mE^2} - \nabla_{\bm{\rho}}^2 + i \, 
 \biggl[ 
   \frac{2\omega_1\omega_2 V^+(\rho)}{\ko\mE^2} 
 \biggr] 
 \biggr\} \, \psi(\bm{\rho}) = 
 0
 \;,
\ee
can be written as
\be
 \psi(\bm{\rho}) = 
 \sum_{\ell = -\infty}^{\infty}
 \frac{u_\ell (\rho)}{\sqrt{\rho}}\, e^{i \ell \phi}
 \;. 
\ee
Among the two solutions for each $\ell$, 
the one regular at origin 
(denoted by $u_\ell^{r} \equiv u^{<}_\ell$) is of the form
\be
 u_\ell^{r}(\rho)
 \, = \, \rho^{1/2 + |\ell|} \, \bigl[ 1 + \rmO(\rho^2) \bigr]
 + i \, \zeta \, \rho^{9/2 + |\ell|} \ln\bigl( \rho/\rho_0 \bigr) + \ldots
 \;, \la{url}
\ee
where $\zeta$ and $\rho_0$ are constants. The coefficient of 
the small-$\rho$ asymptotics of the real part has been fixed 
in a particular way. 
With this normalization, and choosing the solution
regular at infinity as 
$ 
 u_\ell^{>}(\rho) \equiv 
 u_\ell^r(\rho) \int_\rho^\infty \! {\rm d}\rho' / [u_\ell^r(\rho')]^2
$, 
the solution of \eq\nr{full_S_eq} can be written as
\be
 \phi(\bm{\rho},\bm{\rho}') = 
 \sum_{\ell = -\infty}^{\infty} 
 \frac{u_\ell^{<}(\rho)u_\ell^{>}(\rho') e^{i \ell (\phi - \phi')} }
 {2\pi \sqrt{\rho\, \rho'}} 
 \;, \quad
 \mbox{for}\; \rho < \rho' 
 \;. 
\ee
Subsequently we obtain results analogous 
to \eqs(4.25) and (A.33) of ref.~\cite{peskin}:
\ba
 \lim_{\vec{y}\to\vec{0}}
  \frac{\im[g(\vec{y})]}{\pi} & = & 
  \frac{\omega_1\omega_2}{\pi^2 \ko} 
 \int_0^\infty \! {\rm d}\rho \, \im \biggl\{
    \frac{1}{[u_0^{r}(\rho)]^2} 
 \biggr\}
 \;, \la{Swave} \\ 
  \lim_{\vec{y}\to\vec{0}}
  \frac{\im[\nabla_\perp \cdot \vec{f}(\vec{y})]}{\pi}
 & = &
  \frac{4 \omega_1\omega_2\mE^2 }{\pi^2 \ko} 
 \int_0^\infty \! {\rm d}\rho \, \im \biggl\{
    \frac{1}{[u_1^{r}(\rho)]^2} 
 \biggr\}
 \;. \la{Pwave}
\ea
Making use of the symmetry $\omega_1\leftrightarrow\omega_2$
and carrying out one of the integrations, the final expression reads
\ba
 && \hspace*{-1.5cm}
 - \left. \im \Pi^{ }_\rmii{R} \right|^\rmi{full}_\rmii{LPM} 
 \; \equiv \;  
 - \frac{4\Nc}{\pi^2\ko}
 \int_{\ko/2}^{\infty} \! {\rm d}\omega \, 
 \bigl[ 1-\nF{}(\omega)-\nF{}(\ko-\omega) \bigr]
 \nn 
 & \times &  
 \int_0^\infty \! {\rm d}\rho \, 
 \biggl[
   \frac{\omega(\ko-\omega)}{2\ko^2}
   \im \biggl\{ \frac{M^2} 
 {[u_0^r(\rho)]^2} \biggr\} 
 +
   \biggl\{ \frac{\ko^2}{\omega(\ko-\omega)} -2 \biggr\}
   \im \biggl\{ \frac{\mE^2}{[u_1^r(\rho)]^2} \biggr\} 
 \biggr]
 \;, \la{final}
\ea
where the radial wave functions are to be solved from 
\be
 \biggl[
   -\frac{{\rm d}^2}{{\rm d} \rho^2} + \frac{\ell^2-1/4}{\rho^2}
   + \frac{m_\infty^2}{\mE^2} 
   - \frac{\omega(\ko-\omega)}{\ko^2}
     \frac{M^2}{\mE^2} 
   + 2 i\, \frac{\omega (\ko-\omega) V^+}{\ko \mE^2}
 \biggr] \, u_\ell^r(\rho) \; = \; 0
 \;, \la{final_url}
\ee
with the asymptotics 
at $\rho \ll 1$ chosen according to \eq\nr{url}.\footnote{%
 The determination of $u^r_\ell$ and the integration
 over $\rho$ in \eq\nr{final} can be implemented as a simultaneous
 solution of nine real first-order differential equations: for
 $\re u_\ell^r, \im u_\ell^r, \re \partial_\rho u_\ell^r, 
 \im \partial_\rho u_\ell^r$, $\ell = 0,1$,   
 and the $\rho$-integral in \eq\nr{final}. A relative 
 accuracy $\sim 10^{-6}$ can be reached with modest expense
 for all $k,M$ considered. 
 }
As a crosscheck 
we show in appendix~A that \eqs\nr{final}, \nr{final_url}
reduce to the correct free results in the appropriate limit.

%
\section{Combination of the NLO and LPM results}
\la{se:combination}

%
\subsection{Re-expansion of the LPM result and matching with NLO}
\la{ss:expansion}

In order to combine LPM resummation with the NLO result, we need to 
identify those terms in the NLO result which are also part of the LPM
resummation. Care must be taken in order not to count such terms twice. 
The identification can best be carried out by re-expanding the LPM
result as a ``naive'' power series in $g^2$, with the kinematic 
variables $k,M$ treated formally as of $\rmO(\pi T)$, because
this is also the structure inherent to the hard NLO result.   

The gauge coupling appears at two points in \se\ref{ss:basic}:
in the parameter $m_\infty^2$ of \eq\nr{H}, as well as in the potential $V^+$
of \eq\nr{V}. If we expand to zeroth order in $g^2$, \eqs\nr{g_eq}, \nr{f_eq}
can be solved in a Fourier representation.  It is straightforward to check
that \eqs\nr{imPiL_lpm}, \nr{imPiT_lpm} then yield 
\ba
 -\left. \im\Pi^\rmii{ }_\rmii{R,L} \right|^{(g^0)}_\rmii{LPM} & = & 
 \frac{4 \Nc M^2}{\ko^2}
 \biggl[
  2 \bigl\langle \omega(\ko - \omega) \bigr\rangle  
 \biggr]
 \;, \la{rhoL_lo}
 \\ 
 -\left. \im\Pi^\rmii{ }_\rmii{R,T}  \right|^{(g^0)}_\rmii{LPM} & = & 
 \frac{4 \Nc M^2}{\ko^2}
 \biggl[
    \bigl\langle \ko^2 \bigr\rangle 
  - 2 \bigl\langle \omega(\ko - \omega) \bigr\rangle  
 \biggr]
 \;, \la{rhoT_lo}
\ea
respectively, where 
\ba
 \langle ... \rangle 
 & \equiv &  \frac{1}{16\pi \ko}
 \int_{0}^{\ko} \! {\rm d}\omega \, 
 \bigl[
   1  - \nF{}(\omega) - \nF{}(\ko - \omega)   
 \bigr] \, (...) 
 \;. \la{ave_lo}
\ea
A cancellation of 
$
 \bigl\langle \omega(\ko - \omega) \bigr\rangle
$
as discussed in the paragraph following \eq\nr{imPiT_lpm}
is readily verified. 
Summing together
and carrying out the remaining integral, we 
get a limit of \eq\nr{pert_LO}:
\be
  \left. - \im \Pi_\rmii{R}(\mathcal{K}) \right|^{(g^0)}_\rmii{LPM}
 = \frac{ \Nc T M^2 }{2 \pi\ko }
 \; 
 \ln \Bigl[ \cosh \bigl(\frac{\ko}{2 T} \bigr) \Bigr]
 \;. \la{pert_LO_lpm}
\ee
 
We also need the term of $\rmO(g^2)$ from the re-expansion of the LPM
result. Note first that 
the contribution from the potential $V^+$ through \eq\nr{H} is of 
$\rmO(g^4)$ in this counting. One way to see this is that before carrying
out the final integral, the form of the potential is 
\be
  V^{+}_\rmii{ } = 
 \gE^2  \CF  \int \! \frac{ {\rm d}^2\vec{q}}{(2\pi)^2 } \,   
 \Bigl( 1 - e^{i \vec{q}\cdot\vec{y}}\Bigr)
 \biggl( \frac{1}{\vec{q}^2} - \frac{1}{\vec{q}^2 + \mE^2} \biggr)
 \;. 
\ee
We see that if the propagator is expanded to $\rmO(\mE^2)$, 
the term of $\rmO(g^2)$ drops out. 

In contrast, there are two contributions 
of $\rmO(g^2)$
from the mass term $m_\infty^2$. 
Solving \eqs\nr{g_eq} and \nr{f_eq} in a Fourier representation and
taking the cut needed in \eqs\nr{imPiL_lpm} and \nr{imPiT_lpm}, $m_\infty^2$
changes the integration range for the Fourier momentum. In addition, 
it appears explicitly in the integrand, if the Fourier momentum 
originating from $-\nabla_\perp\cdot\vec{f}$ is substituted by other
variables as dictated by the Dirac-$\delta$ constraint from the cut. 
The latter contribution leads to a logarithmic divergence. Determining
the logarithmic term explicitly, and making use of symmetries  
in order to simplify the finite terms, we get
\ba
 & & \hspace*{-1.5cm}
 \left. - \im \Pi_\rmii{R}(\mathcal{K}) \right|^{(g^2)}_\rmii{LPM}
 \; = \;  \frac{ \Nc m_\infty^2 }{4\pi }
\ln \Bigl( \frac{m_\infty^2}{ M^2 } 
    \Bigr)
  \Bigl[ 1 - 2 \nF{}(\ko) \Bigr]
 \nn 
 & & \, + \,  
 \frac{ \Nc m_\infty^2 }{2 \pi }
 \int_0^{\ko} \! {\rm d}\omega \, 
 \bigl[
    \nF{}(\omega) - \nF{}(0) 
  + \nF{}(\ko - \omega) - \nF{}(\ko) 
 \bigr]
 \biggl( \frac{1}{\omega} - \frac{1}{\ko} \biggr)
 \;. \la{pert_NLO_lpm}
\ea
The logarithmic divergence on the first row of \eq\nr{pert_NLO_lpm}
exactly matches that in \eq\nr{log_div}.
For future reference, 
summing together \eqs\nr{pert_LO_lpm} and \nr{pert_NLO_lpm}, we define
\be
 \left. \im \Pi^{ }_\rmii{R} \right|^\rmi{expanded}_\rmii{LPM} 
 \; \equiv \;
 \left. \im \Pi_\rmii{R}(\mathcal{K}) \right|^{(g^0)}_\rmii{LPM}
 \; + \;
 \left. \im \Pi_\rmii{R}(\mathcal{K}) \right|^{(g^2)}_\rmii{LPM}
 \;. \la{lpm_reexp}
\ee

%
\subsection{Numerical evaluation}

The goal now is to combine 
the NLO result from \se\ref{se:hard}
with
the LPM result from \se\ref{se:lpm}. We first
need to subtract from the NLO result those terms that are  
resummed into the LPM expression, 
cf.\ \eq\nr{lpm_reexp}.
After the subtraction, the ``full'' LPM result of \eq\nr{final} can
be added. Thereby the final result reads
\be
 \left. \im \Pi^{ }_\rmii{R} \right|^{ }_\rmi{full} \; \equiv \;
 \left. \im \Pi^{ }_\rmii{R} \right|^{ }_\rmii{NLO}
 \; \underbrace{
 - \; \left. \im \Pi^{ }_\rmii{R} \right|^\rmi{expanded}_\rmii{LPM}
 \; + \; \left. \im \Pi^{ }_\rmii{R} \right|^\rmi{full}_\rmii{LPM}
 \; }_{\equiv 
 \left. \Delta \im \Pi^{ }_\rmii{R} \right|^{ }_\rmii{LPM}
 }
 \;. \la{resummation}
\ee
In a regime where the resummation has no effect, i.e.\
$
 \left. \Delta \im \Pi^{ }_\rmii{R} \right|^{ }_\rmii{LPM}
 = 
 0
$, 
we recover simply the consistent NLO result.
On the other hand, in the soft regime where LPM resummation is important, 
the difference 
$
 \left. \im \Pi^{ }_\rmii{R} \right|^{ }_\rmii{NLO} - 
 \left. \im \Pi^{ }_\rmii{R} \right|^\rmi{expanded}_\rmii{LPM}
$
represents a hard ``matching'' contribution 
(involving $2\leftrightarrow 2$ scatterings and void of the logarithmic
divergence visible in \eq\nr{log_div})
which needs to be added to the soft LPM result. 

Numerical results for 
$
 \left. \Delta \im \Pi^{ }_\rmii{R} \right|^{ }_\rmii{LPM}
$
are shown in \fig\ref{fig:LPM}. It is seen that LPM
resummation has no effect in the hard regime $M \gsim \pi T$.
It does have a substantial 
influence in the regime $M \ll \pi T$, $k \gg M$. 
The behaviour changes qualitatively at small $k$
when the inequality $k \gg M$ 
is no longer satisfied.\footnote{%
 In all regimes, 
 for given $k$ and $M$, $\ko$ was determined from 
 $\ko = \sqrt{k^2 + M^2}$. Then $\ko$ and $M$ were 
 inserted into the expressions of \se\ref{se:lpm}.
 }
However, for any fixed $k > 0$, the curves do 
reach a regime with $k \gg M$ if extrapolated far to the left.
Therefore it is perhaps not completely surprising that
they turn out to be qualitatively correct even
for $k \lsim M$ (cf.\ \fig\ref{fig:NLO}(right)).

\begin{figure}[t]


\centerline{%
 \epsfysize=7.5cm\epsfbox{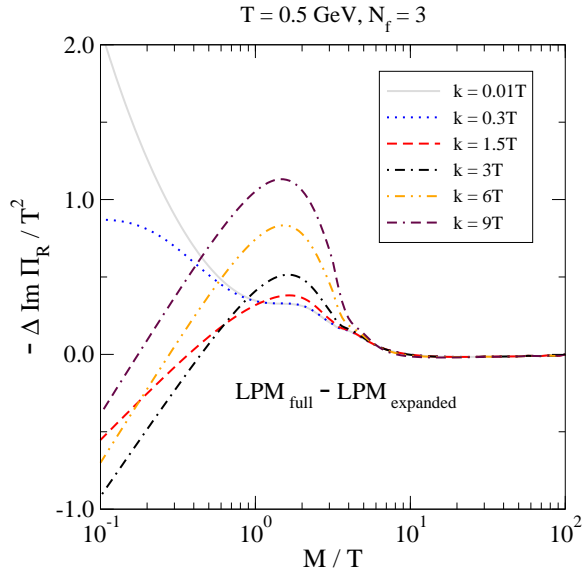}%
}

\caption[a]{\small
The LPM-resummed result, after 
the subtraction of terms already appearing as part of the NLO result. 
It is observed that: (i) In the ``hard'' regime, 
i.e.\ for $M \gsim \pi T$, LPM-resummation has no effect. 
This is because in this regime, 
the loop expansion, and specifically 
the re-expansion of the LPM result, converges rapidly. 
(ii) For small $k$, the behaviour changes qualitatively.
Even though the formal applicability of the result
requires $k\gg \{ gT, M \}$, which is not satisfied in this corner,  
the behaviour nevertheless agrees qualitatively with 
a kinetic theory prediction for $k=0$~\cite{mr}, 
cf.\ \fig\ref{fig:NLO}.  
}

\la{fig:LPM}
\end{figure}

The results obtained after adding 
$
 \left. \Delta \im \Pi^{ }_\rmii{R} \right|^{ }_\rmii{LPM}
$
to the NLO expression 
(\fig\ref{fig:NLO}(left))
are shown in \fig\ref{fig:NLO}(right). 
LPM resummation is seen to remove the logarithmic divergence
of the NLO result at small $M \ll \pi T$  
and leave over a smooth behaviour. 
For very small $k$ the results show an increase which 
is in surprisingly good agreement with 
an effective kinetic theory computation relevant for this regime~\cite{mr}. 

\begin{figure}[t]


\centerline{%
 \epsfysize=7.5cm\epsfbox{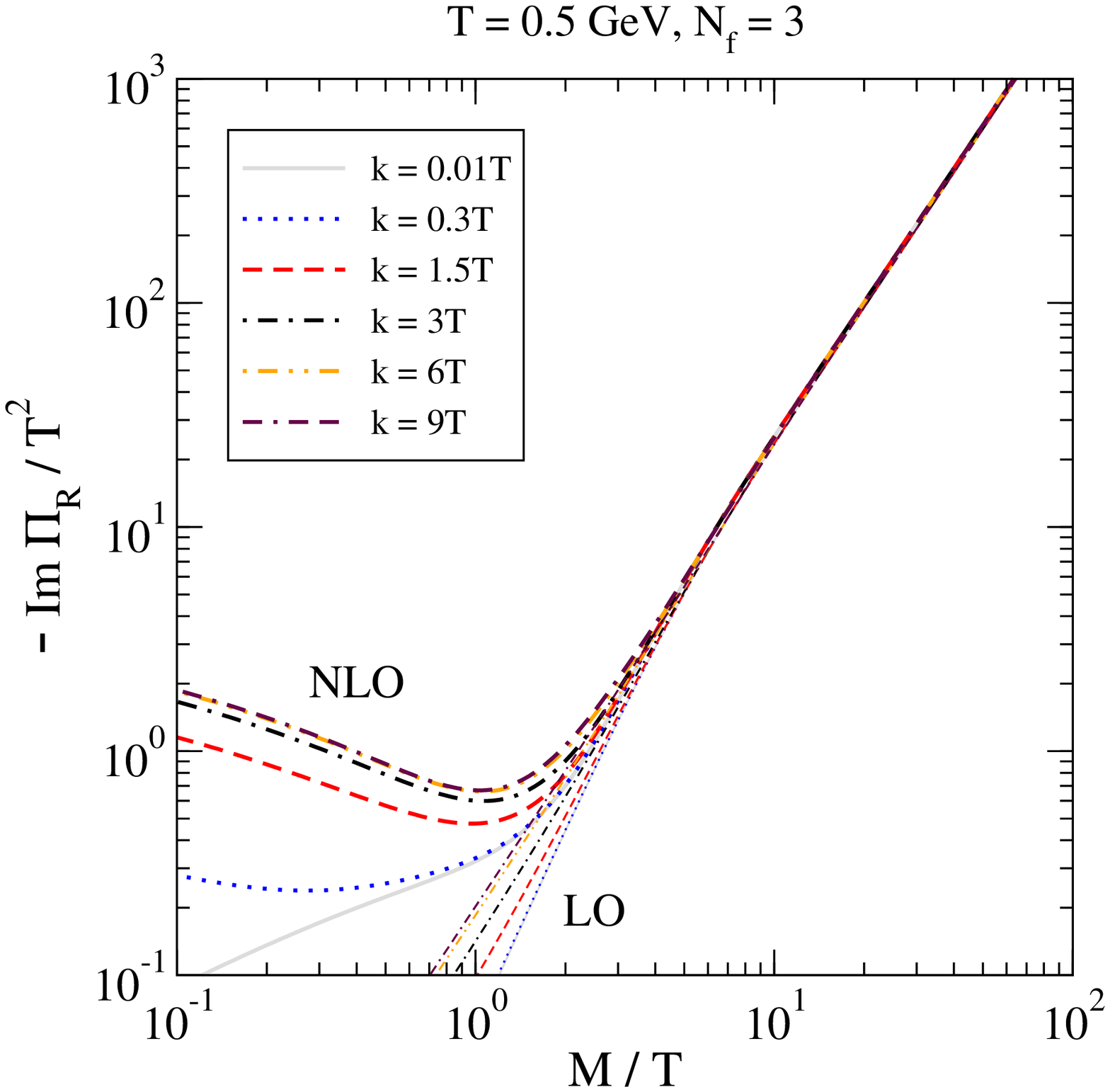}%
~~~\epsfysize=7.5cm\epsfbox{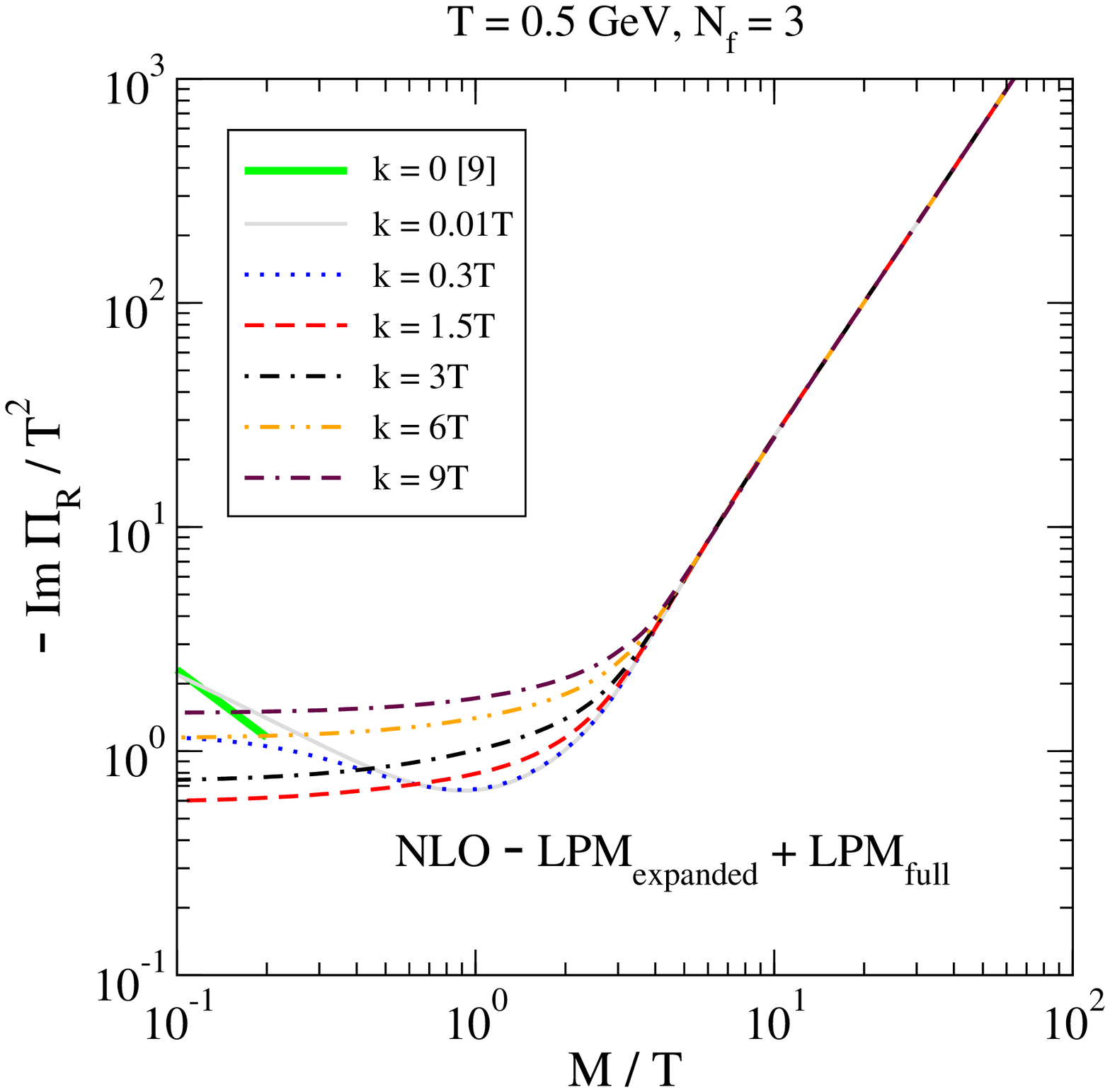}
}

\caption[a]{\small
Left: Strict loop expansion up to NLO, from ref.~\cite{dilepton}.
Right: Results obtained after adding the 
contribution from LPM resummation from \fig\ref{fig:LPM}.   
The renormalization scale has been fixed 
as specified in appendix~B. 
LPM resummation removes the 
logarithmic increase from small $M/T$
and makes the results smoother. 
The correct behaviour for $k=0$ and $g^4 T/\pi^3 \ll M \ll gT$ is 
$- \im \Pi^{ }_\rmii{R} \sim \alpha_s^2 T^3/M$ and has 
been indicated with a green band (from ref.~\cite{mr}, fig.~5).
}

\la{fig:NLO}
\end{figure}

%
\section{Tabulated spectra}
\la{se:spectra}

Given values for $-\im\Pi^{ }_\rmii{R}$, physical dilepton 
rates are given by \eqs\nr{physics}, \nr{relation}. In the 
following we refer to spectra for the production of $\mu^{-}\mu^{+}$
pairs, but the corresponding results for $e^-e^+$ can be obtained by 
a trivial change of the prefactor in \eq\nr{physics}. 
Going over to physical units, {\em viz.}
\be
  \frac{{\rm d} N_{\mu^-\mu^+}}
   {{\rm d}^4 \mathcal{X} {\rm d}^4 \mathcal{K}}
 \times \mbox{GeV}^4\mbox{fm}^4 
 = \frac{{\rm d} N_{\mu^-\mu^+}}
   {{\rm d}^4 \mathcal{X} {\rm d}^4 \mathcal{K}} 
   \biggl( \frac{1000}{197.327} \biggr)^4
  \;, 
\ee 
results are shown for $\Nf = 3$, 
fixing $\Lambdamsbar\simeq 360$~MeV~\cite{pacs}, 
in \fig\ref{fig:rates}. For comparison we display 
both the strict NLO results from ref.~\cite{dilepton} (left panel)
as well as the complete expressions after including LPM 
resummation in the soft regime (right panel).\footnote{%
 The data displayed 
 in \fig\ref{fig:rates} and similar results for 
 other temperatures can be downloaded from
 {www.laine.itp.unibe.ch/dilepton-lpm/}.     \la{fn:table}
 } 
For a fixed 
invariant mass, LPM resummation is seen to have a noticeable
effect at the smallest values of $\ko$, corresponding to 
the smallest spatial momenta $k$. The origin of this 
enhancement can be inferred from \fig\ref{fig:NLO}
(cf.\ the curves $k=0.01T$, $k=0.3T$).

\begin{figure}[t]


\centerline{%
 \epsfysize=7.5cm\epsfbox{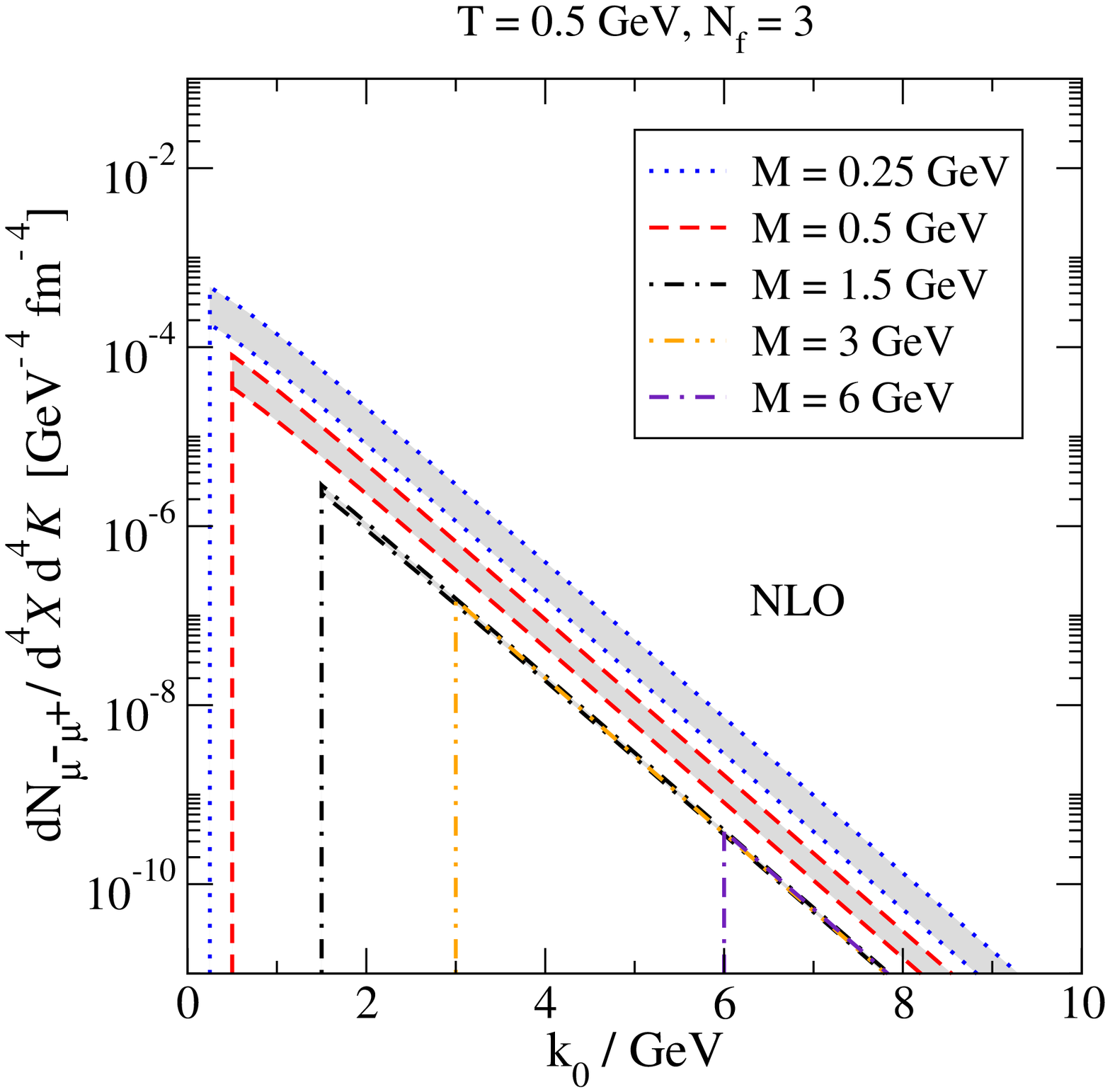}%
~~~\epsfysize=7.5cm\epsfbox{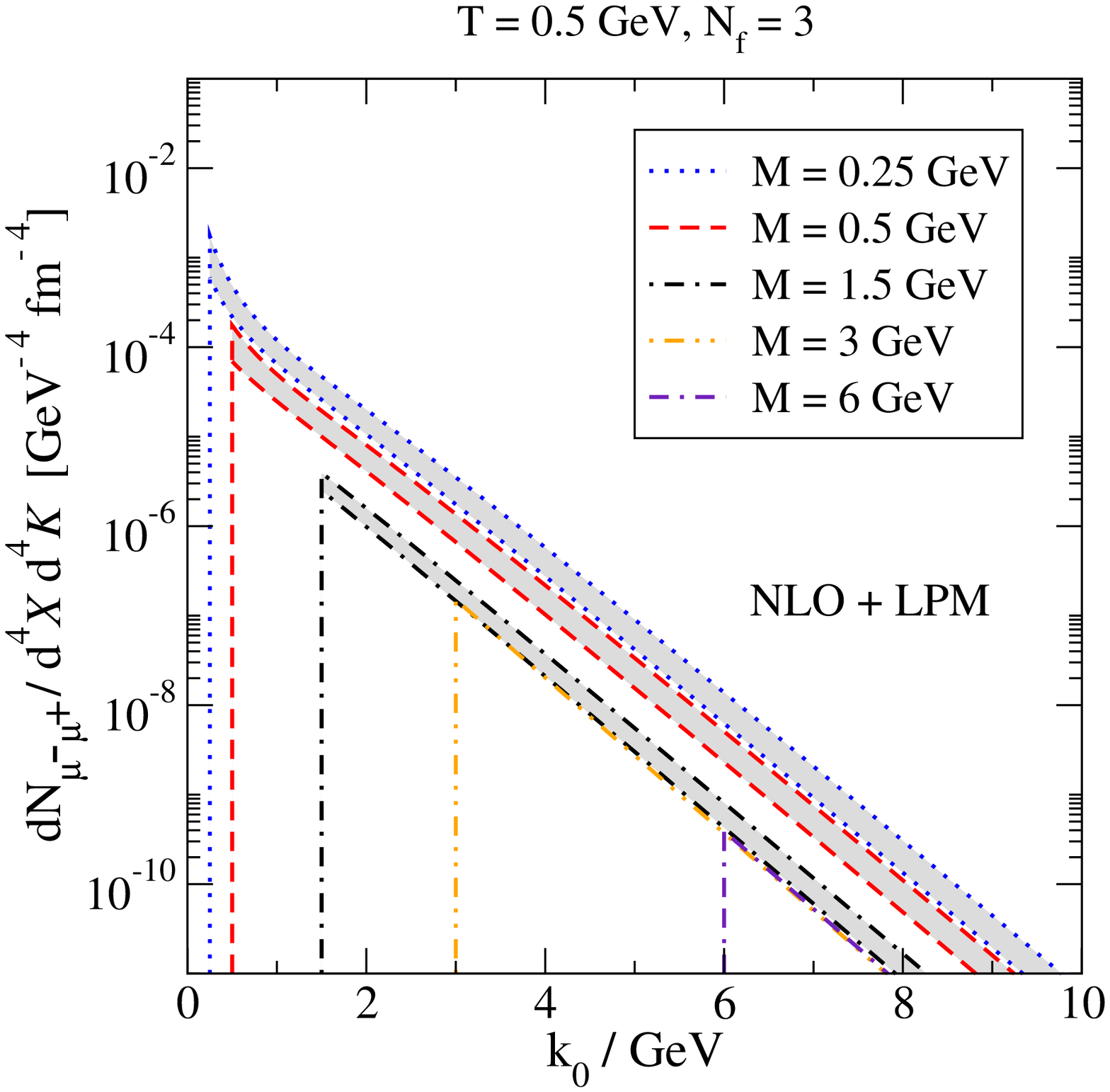}
}

\caption[a]{\small
Left: The NLO dilepton rate, 
for $T = 0.5$~GeV, 
as a function of photon energy~\cite{dilepton}. 
The plots are for $\Nf = 3$
and $\Lambdamsbar = 360$~MeV~\cite{pacs}. 
Bands from scale variation are shown for the three smallest 
photon masses (cf.\ appendix~B).
Right: The same results after adding the contribution
from LPM resummation. The right panel constitutes
our final result at this temperature.
}

\la{fig:rates}
\end{figure}

%
\section{Conclusions and outlook}
\la{se:concl}

The purpose of this paper has been to collect together ingredients from 
two existing computations, namely an LPM-resummed computation
of the thermal dilepton rate in a regime of ``soft'' invariant
masses $M \ll \pi T$~\cite{agz_m},
as well as a full NLO computation in a regime of ``hard'' invariant 
masses $M \gsim \pi T$~\cite{dilepton}. We have shown that the two different
regimes can be ``interpolated'' into a result which 
should represent a fair approximation
(with uncertainties of $\sim 50$\%) for all spatial momenta
and positive invariant masses. The uncertainty estimate is based on
a recent analysis~\cite{screening} 
in which the equations of LPM resummation, analytically continued
to imaginary time, permitted for a determination of
vector channel screening masses
and correlation functions 
which could be compared with 
lattice Monte Carlo data at $T\approx 250$~MeV. 
Our results have been tabulated 
(cf.\ footnote~\ref{fn:table}) 
in a form
which hopefully allows for their insertion into hydrodynamical 
codes such as ref.~\cite{hydro1}. 

{}From the theoretical point of view, 
the results of the present paper are accurate up to NLO ($\rmO(\alphas)$)
for invariant masses $M \gsim \pi T$. For $M \ll \pi T, k \gg M$ they
should still be 
accurate up to LO, thanks to the inclusion of LPM resummation.\footnote{%
 A disclaimer may be in order. 
 In the NLO computation of \se\ref{se:hard}, 
 divergences related to soft momentum 
 transfer cancel between real and virtual corrections. For $k \gg M$, 
 soft momenta are kinematically cut off by a scale 
 $q_\rmii{min} \sim \km \sim
 M^2/(4k)$. For $q_\rmii{min} \ll m_\infty$, 
 i.e.\ $M \ll \sqrt{4 k m_\infty}$,
 this scale is below that at which 
 HTL effects become important. Even though
 infrared contributions cancel 
 even in the presence of HTL effects, 
 which in the NLO computation appear as ``insertions'', 
 it might be questioned whether 
 a supplementary finite term could be left over if the insertions
 were resummed into propagators.
 Excluding explicitly this possibility would require 
 a non-trivial separate computation, 
 which we have not carried out.   
 } 
For $M \ll \pi T$, $k \lsim M$, the results are not  
consistent even at LO, but they nevertheless display a qualitatively
correct behaviour, as a numerical comparison with an effective
kinetic theory analysis~\cite{mr} shows 
(cf.\ \fig\ref{fig:NLO}(right)).  

One way to improve upon our results would be to include NLO corrections of 
$\rmO(\sqrt{\alphas})$ in the soft regime $M \ll \pi T, k \gg M$, 
similarly to what 
has been done for the photon production rate in ref.~\cite{km}.
A systematic study of the very soft regime $M \ll \pi T$, $k \lsim M$
could also be envisaged. Furthermore
the results could in principle be extended into the spacelike domain
$M^2 < 0$, which would allow for another direct comparison with 
lattice measurements, as has been outlined in ref.~\cite{dilepton}.
Finally, it would be interesting to consider non-equilibrium backgrounds, 
probably relevant for practical heavy ion collision experiments. 
We hope to return to some of these challenges in the future. 

%
\section*{Acknowledgements}

M.L acknowledges useful discussions with 
C.~Gale, J.~Ghiglieri and G.~Vujanovic.
This work was partly supported by the Swiss National Science Foundation
(SNF) grant 200021-140234.

%
\appendix
\renewcommand{\thesection}{Appendix~\Alph{section}}
\renewcommand{\thesubsection}{\Alph{section}.\arabic{subsection}}
\renewcommand{\theequation}{\Alph{section}.\arabic{equation}}

%
\section{Free limit of LPM resummation}
\la{app:A}

As a crosscheck, we discuss 
here what happens if the potential $i V^+$ is 
replaced by $i0^+$ in \eq\nr{final_url}.
The correctly normalized regular solutions become
(for $0 < \omega < \ko$)
\ba
 u_0^{r}(\rho) & = & \sqrt{\rho} \, 
 J_0 \Bigl(\rho\sqrt{-M_\rmi{eff}^2/\mE^2 - i0^+}\,\Bigr)
 \;, \la{u0_free} \\ 
 u_1^{r}(\rho) & = & {2 \sqrt{\rho}} \, 
 J_1 \Bigl(\rho\sqrt{-M_\rmi{eff}^2/\mE^2 - i0^+}\,\Bigr)
 \; / \; \sqrt{-M_\rmi{eff}^2/\mE^2 - i0^+}
 \;, 
\ea
and the functions appearing in \eqs\nr{Swave}, \nr{Pwave} read
\ba
 \lim_{\vec{y}\to\vec{0}}
  \frac{\im[g(\vec{y})]}{\pi} & = & 
  \frac{\omega_1\omega_2}{2 \pi \ko} 
 \; \im \left\{ 
 \frac{Y_0\bigl(\rho\sqrt{-M_\rmi{eff}^2/\mE^2 - i0^+} \bigr)}
      {J_0\bigl(\rho\sqrt{-M_\rmi{eff}^2/\mE^2 - i0^+} \bigr)}
 \right\}^{\infty}_{0^+}
 \nn & = &  
 - \frac{\omega_1\omega_2}{2 \pi \ko}
 \,\theta(-M_\rmi{eff}^2)
 \;, \\ 
  \lim_{\vec{y}\to\vec{0}}
  \frac{\im[\nabla_\perp \cdot \vec{f}(\vec{y})]}{\pi}
 & = &
  \frac{4 \omega_1\omega_2 \mE^2}{2 \pi \ko} 
 \; \im \left\{  
 \frac{Y_1\bigl(\rho\sqrt{-M_\rmi{eff}^2/\mE^2 - i0^+} \bigr)}
      {J_1\bigl(\rho\sqrt{-M_\rmi{eff}^2/\mE^2 - i0^+} \bigr)}
 \right\}^{\infty}_{0^+} \biggl( -\frac{M_\rmi{eff}^2}{4 \mE^2} \biggr) 
 \nn & = & 
 - \frac{\omega_1\omega_2}{2 \pi \ko} \bigl( - M_\rmi{eff}^2 \bigr)
 \,\theta(-M_\rmi{eff}^2)
 \;. \la{iu1_free}
\ea
In \eqs\nr{u0_free}--\nr{iu1_free}, $J_0, J_1, Y_0, Y_1$
are Bessel functions of the first and second kind, respectively
($Y_\nu \equiv N_\nu$). 
Inserting these into \eqs\nr{imPiL_lpm}, \nr{imPiT_lpm} and
setting $m_\infty^2\to 0$, the leading-order 
expressions of \eqs\nr{rhoL_lo}--\nr{ave_lo} are reproduced.

%
\section{Choice of parameters}
\la{app:B}

The strong coupling constant runs as 
$
 \partial_t a_s = 
 - (\beta_0 a_s^2 + \beta_1 a_s^3 + \beta_2 a_s^4 + \beta_3 a_s^5 + \ldots)
$, 
where 
$
 a_s \equiv {\alpha_s(\bmu)} / {\pi}
$, 
$
 t \equiv \ln\bigl( {\bmu^2} / {\Lambda_\rmii{$\msbar$}^2} \bigr)
$, 
and, for $\Nc = 3$~\cite{rit1},  
\ba
 \beta_0 & = & \frac{11}{4} - \frac{\Nf}{6}
  \;, \quad 
 \beta_1 \; = \; \frac{51}{8} - \frac{19\Nf}{24}
  \;, \quad 
 \beta_2 \; = \; \frac{2857}{128} - \frac{ 5033\Nf}{1152}
  + \frac{325\Nf^2}{3456}
  \;, \\ 
 \beta_3 & = & \frac{149753 + 21384 \zeta(3) }{1536}
 \nn & - &
      \frac{[1078361+39048\zeta(3)]\Nf}{ 41472 } + 
      \frac{[50065+12944\zeta(3)]\Nf^2}{41472} +
      \frac{1093\Nf^3}{186624}
  \;. 
\ea
The scale parameter $\Lambdamsbar$
represents an integration 
constant and is chosen so that the ultraviolet asymptotics reads
$
  a_s = {1}/({\beta_0 t}) - {\beta_1 \ln (t)}/({\beta_0^3 t^2})
  + \rmO\bigl({1}/{t^3}\bigr)
$.
For numerical results we consider the case $\Nf = 3$
and therefore set $\Lambdamsbar\simeq 360$~MeV~\cite{pacs}.
The renormalization scale is varied within the range 
$
 \bmu \in 
 (0.5 ... 2.0)\, \bmu_\rmi{ref}
$, 
$
 \bmu_\rmi{ref}^2 \equiv {\rm max}\{ \mathcal{K}^2 , (\pi T)^2  \}
$.
In general we have employed 3-loop running 
(i.e.\ $\beta_0, \beta_1, \beta_2$), however we have checked that
results obtained with 4-loop running are well within the error band.

%

\end{document}